\documentclass[floatfix,twocolumn,aps,prl,showpacs,amsmath,amssymb]{revtex4-1}
\usepackage{epsfig}
\usepackage[latin1]{inputenc}

\usepackage[dvips]{color}

\newcommand{\n}{\noindent}

\begin{document}

\title{ Interaction Induced Quantum Valley Hall Effect in Graphene}

\author{E. C. Marino$^1$, Leandro O. Nascimento$^{1,2}$, Van S\'ergio Alves$^{1,3}$, and C. Morais Smith$^2$}
\affiliation{$^1$Instituto de F\'\i sica, Universidade Federal do Rio de Janeiro, C.P.68528, Rio de Janeiro RJ, 21941-972, Brazil \\
$^2$Institute for Theoretical Physics, Centre for Extreme Matter and Emergent Phenomena, Utrecht University, Leuvenlaan 4, 3584CE Utrecht, the Netherlands \\
$^3$Faculdade de F\'\i sica, Universidade Federal do Par\'a, Av.~Augusto Correa 01, 66075-110, Bel\'em, Par\'a, Brazil }

\date{\today}

\begin{abstract}
 
We use Pseudo Quantum Electrodynamics (PQED) in order to describe the full electromagnetic interaction
of the p-electrons of graphene in a consistent 2D formulation. We first consider the effect of this interaction
in the vacuum polarization tensor or, equivalently, in the current correlator. This allows us to obtain the dc 
conductivity after a smooth zero-frequency limit is taken in Kubo's formula.Thereby, we obtain the usual expression for the 
minimal conductivity plus corrections due to the interaction that bring it closer to the experimental value.
 We then predict the onset of an interaction-driven spontaneous Quantum Valley Hall effect (QVHE) below a critical 
temperature of the order of  $0.05$ K. The transverse (Hall) valley conductivity is evaluated exactly and shown to coincide
with the one in the usual Quantum Hall effect. Finally, by considering the effects of PQED, we show that the electron self-energy
is such that a set of P- and T- symmetric gapped electron energy eigenstates are dynamically generated, in association
with the QVHE.
 
\end{abstract}


\maketitle

{\it Introduction.--}
The experimental realization of graphene has opened the fascinating possibility of observing in a condensed matter system
 a number of interesting effects previously considered to occur exclusively in relativistic particle physics. The Klein paradox \cite{Katsnelson} and the Zitterbewegung \cite{Katsnelson1} are well-known examples. Graphene is also the first concrete realization of the Dirac sea, the concept which has led Dirac to predict the existence of antimatter. Indeed, Schwinger's effect of pair creation out of the vacuum by an electric field is expected to occur in this material, thus providing another beautiful connection between condensed matter and particle physics \cite{Allor}.

Graphene exhibits quite a few unconventional transport phenomena. These include an anomalous integer quantum Hall effect (QHE) \cite{GeimIQHE} and a puzzling finite (``minimal'') dc conductivity at half-filling \cite{opticalexpt}, even in the absence of any dissipation and with a zero density of states. The theoretical determination of the minimal dc conductivity and its dependence on interactions is still a challenge \cite{condmin}, in part due to the ambiguities associated to the $\omega \rightarrow 0$ limit in Kubo's formula. Attempts to include the effect of interactions in the calculation of the optical conductivity were recently made \cite{Juricic}, however this effect disappears in the limit $\omega \to 0$ and therefore no corrections to  $\sigma_{dc} $ due to the interactions can be obtained.

Nevertheless, optical conductivity measurements \cite{opticalexpt}  yielded results that in the dc limit are in agreement with earlier theoretical calculations in the approximation of non-interacting electrons, namely $\sigma_{dc} = (\pi/2) e^2/h$ \cite{theorynonint}. Analogously, the integer QHE \cite{GeimIQHE} has been understood in terms of relativistic Landau levels occupied by non-interacting electrons, similarly to the results for GaAs \cite{Mark,review}.

The unexpected validity of the single-particle description has risen the issue of how relevant are the electronic interactions in graphene, leading to a vivid debate in the community. Nonetheless, the recent measurement \cite{Elias} of the renormalization of the Fermi velocity \cite{Vozmediano} was an indication that interaction should be important. The direct measurement of the dc conductivity \cite{Andrei2}, which yielded a result that is in disagreement with the theoretical calculation in the absence of interactions, provided additional evidence for the relevance of these. The experimental observation of the fractional QHE in ultra-clean samples subject to a perpendicular magnetic field has closed the debate, undeniably demonstrating that the electronic interactions are indeed important, at least for a certain energy (temperature) scale \cite{Andrei,Kim,Kim11,Kim11a}.

Another intriguing transport property that has been investigated in graphene is the possibility of observing a quantized transverse (Hall) conductivity under unconventional circumstances. First, Haldane has shown that the sufficient condition for the existence of the integer QHE is a broken time-reversal symmetry (TRS) and not a net magnetic field, as was previously supposed \cite{Haldane}. Later, even more unexpected results emerged, such as the experimental observation of the integer QHE {\it at} {\it room} {\it temperature} \cite{Nijmegen} and the proposal for the existence of a quantum {\it spin} Hall effect in the presence of a sizable spin-orbit coupling in a system which preserves time-reversal symmetry (TRS) \cite{Kane}.

Most of the previous approaches, however, rely on a single-particle description and the role of interactions, as well as the proper theoretical framework to include them, has been often neglected or only partially included. From first principles, the relevant electronic interaction in graphene is the full electromagnetic interaction described by the minimal coupling of the electronic current to the U(1) electromagnetic gauge field. This, however, is not easily incorporated in the model because the electrons in graphene are confined to a plane and therefore require a 2D description, whereas the electromagnetic field is 3D. Should we use Maxwell electrodynamics in 2D for describing the interaction of the electrons in graphene, we would get a wrong result (for instance the electrostatic potential would be $-\ln r$ instead of the correct $1/r$). The solution for this problem consists in the use of a full 2D U(1) gauge field theory, which describes, within the 2D framework, the full physics contained in the 3D Maxwell theory. Such 2D theory was derived in \cite{Kovner} in the static limit. Subsequently a full dynamical derivation was provided \cite{marino} and the theory was called Pseudo Quantum Electrodynamics (PQED) (in part because it envolves the so-called pseudo-differential operators) \cite{marino,marinorubens,marino1} .

In this letter, we employ PQED in order to describe the electronic interactions in graphene and explore some of the consequences of these. We firstly determine the corrections to the minimal dc-conductivity produced by such interactions, thus obtaining a value at $T=0$, which is the closest to the measured experimental one \cite{Andrei2}. We then evaluate the effects of PQED in the valley dc-conductivity and show  that, below a temperature $T_c$, it exhibits a nonzero transverse component, which is quantized in the same way as in the usual QHE. This effect is {\it dynamically} generated in graphene, when the {\it full} electromagnetic  interaction is {\it completely} taken into account. In this case, we show that the individual valley contribution to the conductivity contains a P,T-violating transverse (Hall) component, which has opposite sign for each valley and consequently leads to a Quantum Valley Hall Effect (QVHE), rather than the usual QHE. Finally, we investigate the Schwinger-Dyson equation for the electron self-energy in PQED and show that the latter satisfies a differential equation, which has solutions that shift the poles of the electron propagator to gapped energy states, when the interaction coupling is larger than a certain critical value. The temperature scale is set by the gap: thermal activation will destroy the plateaus for temperatures larger than the gap.  

All the phenomena described here only occur within an SU(2) description of graphene, which is valid when there is no backscattering connecting the different valleys. We also show that the use of PQED, contrary to other attempts to describe the electronic interactions in graphene, yields a current correlator that renders Kubo's formula free from any ambiguities. In the last section, we show that our results are independent of the fact that the Fermi velocity is different from the speed of light.

{\it The} {\it Model. --}
The $p$-electrons of the carbon atoms in the honeycomb lattice of graphene are usually described as 4-component massless Dirac fermions, each component corresponding to the two sublattices (A and B) and the two inequivalent valleys (K and K'). If we neglect backscattering between the valleys, however, an equivalent description would consist of two massless 2-component Dirac fermion fields.
Backscattering involves momenta of order $\hbar/a$, hence this process must be triggered by lattice displacements of order $a$, which in the quantum version are phonons. In any description that does not consider phonon effects, actually one should, for consistency, neglect backscattering. We assume that these Dirac electrons will interact through the electromagnetic interaction, which in 2D is described by  PQED \cite{marino}. The corresponding Lagrangian reads
\begin{eqnarray}
{\cal L}&=& \frac{1}{4} F_{\mu \nu}\left[\frac{4}{\sqrt{-\Box}}\right] F^{\mu\nu} + \bar\psi_a\left(i\partial\!\!\!/+e\,\gamma^{\mu}\, A_{\mu}\right)\,\psi_a,
\label{action}
\end{eqnarray}
where $\psi$ is a two-component Dirac field, $\bar\psi =\psi^\dagger\gamma^0$ is its adjoint, $F_{\mu \nu}$ is the usual field intensity tensor of the U(1) gauge field $A_\mu$, which intermediates the electromagnetic interaction in 2D (pseudo electromagnetic field),  $\gamma^\mu$ are rank-2 Dirac matrices, and $a=1,...,N_f$ is a flavor index, specifying the spin component and the valley to which the electron belongs. An SU(4) version of this model has been recently used to study dynamical gap generation and chiral symmetry breaking in graphene \cite{VLWJF}.

A few remarks are in order here: {\it i)} the natural velocity appearing in the gauge field sector is that of light, $c$,
whereas the one occurring in the electronic sector is the Fermi velocity $v_{\rm{F}}$, hence Lorentz invariance
is broken. For the time being we shall take $v_{\rm{F}} = c = 1$. Later on we shall return to the physical
value of the Fermi velocity; {\it ii)} the gauge field propagator in momentum space contains the speed of light  $c$ rather
than $v_{\rm{F}}$, hence there will be no retardation effects due to the fact that $v_{\rm{F}} \sim c/300; $ {\it iii)} the two valleys (K and K') are related by TRS. In spite of the fact that we are using 2-component Dirac spinors, however, we do not break TRS {\it ab initio} because we are summing over the two species (as we may infer from the physical value of $N_f$).

{\it The current-current correlation function.--}
We are going to determine the
dc conductivity of graphene by using the Kubo formula, which requires the evaluation of the irreducible two-points current correlation function
 $\langle j^\mu j^\nu \rangle$.
This is most conveniently obtained as a second functional derivative of the generating functional of 1PI Green function $\Gamma [A^{\mu}_c]$,
\begin{eqnarray}\label{correlation}
\langle j_\mu j_\nu \rangle_{1PI}=\frac{1}{e^2}\frac{\delta^2}{\delta A_c^{\mu} \delta A_c^{\nu}}\Gamma [A_c^\mu]\Big| _{A_c^\mu=0}=\Pi_{\mu\nu}, \label{jj}
\end{eqnarray}
where $\Pi_{\mu\nu}$ is the full vacuum polarization tensor.

For two-component fermions this tensor was carefully calculated in Ref.~\cite{Luscher}. The one-loop result for a single massless fermion in Euclidean space is
\begin{eqnarray}
\Pi_{\mu\nu}^{(1)}(p)= -\frac{\sqrt{p^2}}{16} P_{\mu\nu}+\frac{1}{2\pi}\left(n+\frac{1}{2}\right)\epsilon_{\mu\alpha\nu}p^\alpha,
\label{pi}
\end{eqnarray}
where $n$ is an integer (see Supplemental Material). 
Even though this result was derived for QED3, it also holds for PQED because it only involves fermion internal lines.
Notice the occurrence of a P,T-breaking term, which is topological and according to the Coleman-Hill theorem \cite{ch} has no higher order
corrections.

The two-loops correction is exclusive of PQED and was calculated in Ref. \cite{Teber} for a single massless fermion, yielding
\begin{equation}
\Pi_{\mu\nu}^{(2)}(p)=-\frac{\sqrt{p^2}}{16}\,C_{\alpha}\,\alpha_g\,P_{\mu\nu}\,, \label{pi2}
\end{equation}
where $C_{\alpha}=(92-9\pi^2)/18\pi\approx 0.056$ and $\alpha_g\approx 300/137 = 2.189$. 
It follows that $C_{\alpha}\,\alpha_g <1$, hence the perturbation expansion is justified.

{\it The dc conductivity $\sigma^{ij}$.--}
The dc conductivity can be derived, within the linear response regime, from Kubo's formula, which for real time is:
\begin{eqnarray}
\sigma^{ik}= \lim_{\omega\rightarrow 0, \textbf{p}\rightarrow 0}\frac{i\ \langle j^i j^k \rangle_{ret.}}{\omega} =\sigma_{xx}\delta^{ik}+\sigma_{xy} \epsilon^{ik0}.
\label{kubo}
\end{eqnarray}

However, we must sum over the spin components and over the valleys K and K'. The sum over spins just contributes a
factor $2$ to $N_f$. Since the P,T-symmetries are preserved, the contributions from the two valleys are clearly identical and the conductivity reads 
\begin{equation}
\sigma^{ik}= \lim_{\omega\rightarrow 0, \textbf{p}\rightarrow 0}\left\{\frac{i \langle j^i j^k \rangle}{ \omega}
+ \frac{i \langle j^i j^k \rangle^T}{ \omega} \right\},
\label{kubo1}
\end{equation}
where $\langle j \ j \rangle^T$ is the time-reversed correlator and, before summing on the  valleys
we are using $N_f =2$ (see the Supplemental Material for more details). Using Eq.(\ref{pi}) and Eq.(\ref{pi2}) and taking the limits in Eq.(\ref{kubo1}) we get
\begin{equation}
\sigma_{xx}= \left(\frac{\pi}{2} \frac{e^2}{h}\right)\left[1+\left(\frac{92-9\pi^2}{18\pi}\right)\,\alpha_g+{\cal O}(e^4)\right]
\label{imp11}
\end{equation}
and
\begin{equation}
\sigma_{xy}= 0.
\label{imp22}
\end{equation}

Eq. (\ref{imp11}) displays the dc conductivity, which has the usual minimal value plus corrections due to the interaction.
We emphasize that the corrections above are to the {\it dc conductivity}, rather than to the optical conductivity $\sigma (\omega)$ usually found in the literature \cite{Mastro}. To the best of our knowledge, the value we find for the dc conductivity, namely $\sigma_{xx} = 1.76$
$e^2/h$ is the closest to the experimental result extrapolated to zero temperature,  $\sigma_{xx} = 2.16$ $e^2/h$ \cite{Andrei2}.

{\it Quantum Valley Hall Effect. --} The average valley current is defined by
\begin{equation}
\langle J^i_V \rangle = \langle 0|j^i_K|0\rangle - \langle 0|j^i_{K^{\prime}}|0\rangle.
\end{equation}

We can therefore define a dc ``valley conductivity'', through
\begin{equation}
\sigma^{ik}_V= \lim_{\omega\rightarrow 0, \textbf{p}\rightarrow 0}\left\{\frac{i \langle j^i j^k \rangle}{ \omega}
- \frac{i \langle j^i j^k \rangle^T}{ \omega} \right\},
\label{kubovsup}
\end{equation}
where $\langle j \ j \rangle^T$ is the time-reversed  correlator and the sum over
spins is assumed to have been done. One immediately concludes that, for $T<T_c$, the  valley conductivity, is given by (see Supplemental Material)
\begin{equation}
\sigma^{xy}_V=4\left(n+\frac{1}{2}\right)\frac{e^2}{h},
\label{svxy}
\end{equation}
for $n=$ integer. The longitudinal component, conversely, vanishes:
\begin{equation}
\sigma^{xx}_V= 0.
\end{equation}
The above result is exact, as a consequence of the Coleman-Hill theorem. The existence of a transverse valley conductivity characterizes the occurrence of a QVHE. It is caused, ultimately, by the presence of the anomalous P,T violating term appearing in the vacuum polarization tensor or, equivalently, in the current  correlator.

The valley Hall effect has been earlier predicted to occur in graphene systems subject to a staggered sublattice potential that breaks inversion symmetry  \cite{Xiao,Yao}, or to strained graphene, where according to recent experiments pseudomagnetic fields oppositely oriented in the valleys can be as large as 300 T  \cite{Crommie}. In addition, a fractional valley Hall effect was proposed to arise in artificial graphene systems, by fine tuning the short-range part of interactions \cite{Ghaemi}.  Notice that here no symmetry is broken a priori and no fine-tuning of model parameters are required to generate the QVHE. A similar TRS breaking was recently proposed to occur for bilayer graphene in the presence of static Coulomb interactions, when fluctuations are taken into account \cite{Nandkishore}.

The anomalous terms found here are related to electron masses that are dynamically generated. These, however, appear in
pairs of opposite signs and, for an even number of flavors, cancel when summed, according to the Vafa-Witten theorem \cite{VW}.
There is, consequently no overall P, T violation and for this
reason the QHE does not occur. The existence of individually violating terms, nevertheless, is
sufficient to produce a QVHE. This is the central result of this work.

{\it Dynamically generated discrete energy states and $T_c$.--}
Recently, it has been shown that the model described by Eq.~(\ref{action}) dynamically generates
a gap in the SU(4) case due to a breaking of the chiral symmetry \cite{VLWJF}. The result is obtained by a non-perturbative solution of
the Schwinger-Dyson equation \cite{Robert}
\begin{equation}
S^{-1}_F(p)=S_{0F}^{-1}(p)-\Sigma (p),
\label{sd}
\end{equation}
where $S_{0F}$ and $S_{F}$ are, respectively, the free and interacting electron propagators and $\Sigma(p)$, the electron self-energy calculated in PQED \cite{VLWJF}.
Here, we investigate the SU(2) case and show that an infinite sequence of discrete
energy eigenstates is dynamically generated.  For the two-component Dirac fields considered here, the associated gap generation breaks the parity and TRS symmetries 
instead of the chiral one. The generation of this set of eigenstates, which according to Eq.~(\ref{sd}) is a consequence
of the interactions, is therefore another manifestation of the dynamical symmetry breaking found in the vacuum polarization tensor,
which has led to the spontaneous QHE for each valley below $T_c$.

By imposing the eigenenergy ($\epsilon$) to satisfy
\begin{equation}
\Sigma (p=\epsilon)=\epsilon ,
\label{ae}
\end{equation}
we guarantee, according to Eq.~(\ref{sd}), that they will be zero-momentum poles of the complete electron propagator, thus
renormalizing the tight-binding eigenenergy $\epsilon = |\vec k|$.
The energy levels are discrete and given by
\begin{eqnarray}\label{en}
\epsilon^{(\pm)}_n=\pm\Lambda\, \exp\left\{-\frac{Z_n}{\gamma}\right\},
\end{eqnarray}
where
\begin{equation}
\gamma= \frac{1}{2}\sqrt{\frac{N_c}{N_f}-1}, \qquad
N_c =\frac{4\lambda}{\pi^2 (4+\frac{\lambda}{16})},
\label{N}
\end{equation}
$\Lambda \sim \pi/a$ is an ultraviolet cutoff, $n=0, 1, 2,...$,
and $Z_n =n \pi+  \delta_n$ are solutions of  the transcendental equation
\begin{equation}
\exp\left(-\frac{3}{2\gamma}z\right) =  \cos z,
\label{ee1}
\end{equation}
such that
\begin{eqnarray}
&\ &0 \leq \delta_n \leq \frac{\pi}{2} \ \ \ \ \  n=0, 2, 4, ... \nonumber \\
&\ & \frac{\pi}{2} \leq \delta_n \leq \pi \ \ \ \ \  n=1, 3, 5, ...
\end{eqnarray}
Inversion of Eq.~(\ref{N}) allows us to determine a critical interaction $\lambda_c$ above which the phenomena described here will occur. 

For all $n$, $\delta_n \rightarrow \pi/2 $ for $n \rightarrow \infty$,
whereas  $\delta_n \rightarrow 0 $ ($n$ even) and $\delta_n \rightarrow\pi$ ($n$ odd) for $\gamma \rightarrow \infty$
(an unphysical limit, in which $N_f \rightarrow 0$, see Supplemental Material). Since $\gamma$ is small, $Z_n \approx
(2 n + 1 ) \pi /2$.

Notice that $\gamma$ is real for $N_f<N_c$ and all the $\epsilon_n$ collapse to zero as $N_f\rightarrow N_c$.
Observe also that  the sequence of negative levels labeled by $n=0, 1, 2,...$ have increasing energies and accumulate at
zero for  $n \rightarrow \infty$, whereas the positive levels labeled by $n=0, 1, 2,...$ have decreasing energies and accumulate at
zero for  $n \rightarrow \infty$, being therefore symmetric with respect to zero.

We finally remark that the existence of a one-to-one mapping between the energy bands with the respective gaps and the valley Hall conductivity plateaus, which count the number of edge states, is simply a manifestation of the bulk-boundary correspondence, known to apply for topological insulators.

We may now estimate the  critical temperature $T_c$ for the observation of the effect we found. This must correspond to the
thermal activation energy, which is of the order of the maximal gap $\Delta_\infty = \epsilon^{(+)}_0-\epsilon_\infty=\epsilon_\infty-\epsilon^{(-)}_0 $
(this is of the same order as the first gap $\Delta_1=|\epsilon^{(\pm)}_0|-|\epsilon^{(\pm)}_1| \approx \Delta_\infty $, especially for small $\gamma$).

Supposing that a fraction $1-x$ of the electrons in the ground state are promoted to higher levels by thermal activation,
we would have (see Supplemental Material)
\begin{equation}
T_c \simeq  \frac{\Lambda}{k_B x}  \exp\left[-\frac{\pi}{2\gamma}\right],
\end{equation}
\n where we used Eq.~(\ref{en}) and the fact that $\delta_0 \simeq \pi/2$.
Observe that $T_c \rightarrow 0$ as $N_f\rightarrow N_c$.

In Fig.1, we plot $T_c$ as a function of
the coupling
$\alpha$ for $N_f = 4$ and $x=0.01$. We estimate the upper temperature threshold for observation
of the spontaneous QVHE in suspended graphene in the vacuum ($\alpha_g \approx 2.189$) to be of the order of $0.05$ K. Note that we have used the physical value of the Fermi velocity in the expression of the fine-structure constant of graphene to plot Fig. 1.

\begin{figure}[htb]
\label{criticalT3}
\centering
\includegraphics[scale=0.6]{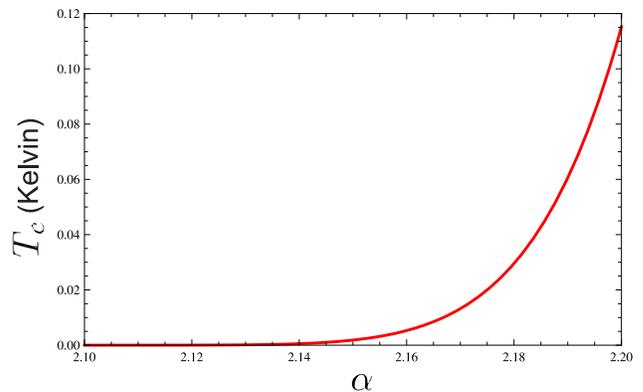}
\caption{(Color online) Critical temperature $T_c$ as a function of the coupling
$\alpha$ for $N_f = 4$,  $x=0.01$ and $\Lambda \sim \hbar v_{\rm{F}}/a$. }
\end{figure}

{\it Fermi velocity.--}
Let us discuss now the consequences of the fact that the Fermi velocity is different from $c$. For describing
this effect, one must make the replacements $\gamma^i \rightarrow \gamma^i v_{\rm{F}}$ in the Dirac kinetic term. Because of the
linear dependence of the Dirac Lagrangian on $p = (p_0,{\bf p})$, it follows that all dependence on $v_{\rm{F}}$ will
appear in the form $v_F {\bf p}$.
In an analogous way, the current will change as
   $j_{\mu}=(j_0,j_i)  \rightarrow j_\mu=(j_0,v_{\rm{F}} j_i)$ and the current correlation function we used in Eq.~(\ref{kubo}) is actually  $v_{\rm{F}}^2\langle jj\rangle$.

The whole dependence of the conductivity on $v_{\rm{F}}$ comes through the insertion of the current correlator 
in the Kubo formula, after the corresponding limits are taken. The current correlator dependence on the Fermi velocity,
on its turn, comes through the vacuum polarization tensor, as we may infer from Eqs.~(\ref{jj}). Now, the different
 components of $\Pi_{\mu\nu}$ depend on $v_{\rm{F}}$ only through the combination $v_F {\bf p}$. When taking the limit
 $ {\bf p} \rightarrow 0$ in Kubo's formula, therefore, any dependence on $v_F$ shall completely disappear (see Supplemental Material).

The net effect of reinstating the physical values of $v_{\rm{F}}$ and $c$ in the self-energy can be shown to be a rescaling of the rhs of Eq. (\ref{ee1}) by a dimensionless factor $f_1(v_{\rm{F}},c)$. Yet, since $\gamma$ is small the lhs is practically equal to zero and consequently the zeros of the transcendental equation coincide with the ones of the $\cos z$ function irrespective of the value of $f_1$.

{\it Summary.--}
Experimental and theoretical results suggest that electronic interactions must be important in graphene, at least for  a certain temperature range. The observation of the fractional QHE \cite{Andrei,Kim,Kim11} is an example of the former, whereas renormalization group calculations, which show an increase of the interaction strength as we lower the temperature \cite{Vozmediano}, is an example of the latter. We have provided a complete and strictly 2D description of the real electromagnetic interactions occurring among the electrons in graphene. For the longitudinal dc conductivity we obtain the ``minimal'' value plus corrections due to the interaction, which make it closer to the experimental result. In addition, the interaction generates a dynamical TRS breaking through one-loop vacuum fluctuations. This produces, below a critical temperature $T_c$, a transverse (Hall) valley conductivity quantized exactly as if there was an external magnetic field in the QHE. Discrete states corresponding to the Hall plateaus, and analogous to the Landau levels in the usual QHE, appear as interaction induced renormalized poles of the fully corrected electron propagator at zero momentum. 

The quantization of our valley currents is {\it emergent,}  {\it exact} and {\it universal}, contrarily to the results obtained in the literature for a QVHE driven by inversion symmetry breaking (staggered chemical potential) \cite{Xiao,Yao,Ezawa}. 
Even though our calculations are made at $T=0$, we may estimate the critical temperature for observing the effect by identifying $T_c$ with the gap. This follows from the  fact that, when the temperature reaches this level, most of the states would be populated by thermal activation and, thereby, the plateaus would be washed out by thermal activation. 

The electrons from different valleys are sensitive to the circular polarization of light \cite{Ezawa}, thus producing a circular dichroism whenever they are spatially separated. Scattering of unpolarized light at $T\approx 0.05$ K, therefore, should be an experimental way for observing this fascinating effect.

\acknowledgments

This work was supported in part by CNPq (Brazil), CAPES (Brazil), FAPERJ (Brazil), the Netherlands Organization for Scientific Research (NWO) and by the Brazilian
government project Science Without Borders.
We are grateful to G.'t Hooft,  A. H. Castro Neto, M. Goerbig, L. Fritz, and V. Juri\v ci\' c for interesting discussions.

\newpage
{\bf Supplemental Material for ``Interaction Induced Quantum Valley Hall Effect in Graphene"}

E. C. Marino$^1$, Leandro O. Nascimento$^{1,2}$, Van S\'ergio Alves$^{1,3}$, and C. Morais Smith$^2$ \\
$^1$Instituto de F\'\i sica, Universidade Federal do Rio de Janeiro, C.P. 68528, Rio de Janeiro RJ, 21941-972, Brazil \\
$^2$Institute for Theoretical Physics, Centre for Extreme Matter and Emergent Phenomena, Utrecht University, Leuvenlaan 4, 3584CE Utrecht, the Netherlands \\
$^3$Faculdade de F\'\i sica, Universidade Federal do Par\'a, Av.~Augusto Correa 01, 66075-110, Bel\'em, Par\'a, Brazil 

\section{The current-current correlation function}

The accurate description of the electromagnetic interaction among the electrons in graphene is given by the following Lagrangian
\begin{eqnarray}
{\cal L}= \frac{1}{4} F_{\mu \nu}\left[\frac{4}{\sqrt{-\Box}}\right] F^{\mu\nu} + \bar\psi_a\left(i\partial\!\!\!/+e\,\gamma^{\mu}\, A_{\mu}\right)\,\psi_a.
\label{actionsup}
\end{eqnarray}
Here, $\psi$ is a two-component Dirac field, $\bar\psi = \psi^\dagger \gamma^0$ is its adjoint, $F_{\mu \nu}$ is the usual field intensity tensor of the U(1) gauge field $A_\mu$, which intermediates the electromagnetic interaction in 2D (pseudo electromagnetic field),  $\gamma^\mu$ are rank-2 Dirac matrices, $a=1,...,N_f$ is a flavor index, specifying the spin component and  the valley to which the electron belongs. The coupling constant $e^2= 4\pi\alpha$ is conveniently written in terms of $\alpha$, the fine-structure constant in natural units.
The Lagrangian described by Eq.~(\ref{actionsup}) was derived within the pseudo-QED formalism, which appropriately takes into account the fact that the electromagnetic field is 3D, whereas the dynamics of electrons in graphene is 2D \cite{marinosup1,marinosup2,marinorubenssup,Kovner}.

We will determine the
dc conductivity of graphene by using the Kubo formula, which describes the linear response to a static external eletric field. 
In real time, it is given by
\begin{equation}
\sigma^{ik}= \lim_{\omega\rightarrow 0, \textbf{p}\rightarrow 0}\frac{i \langle j^i j^k \rangle}{\omega}.
\label{kubosup}
\end{equation}
where the current correlation function is meant to contain only one-particle-irreducible (1PI) diagrams \cite{mahan}.

The current correlator is most conveniently obtained from  the corresponding generating functional.
Starting from the generating functional of arbitrary correlators,
\begin{eqnarray} \label{partitionsup}
{\cal Z}[J]&=&{\cal N}\int DA_\mu D\bar\psi D\psi  \, {\rm e}^{-\int d^3x({\cal L}+e\, j^\mu J_\mu)}\,,
\end{eqnarray}
where $J$ is a vector functional variable and ${\cal N}={\cal N}_{A}{\cal N}_{\psi}$ are constants chosen in such a way that ${\cal Z}[0]=1$ ,
we have that the generator of connected correlation functions is given by
\begin{equation}
W[J]=-\ln {\cal Z}[J]\, .
\end{equation}

The generating functional of 1PI correlation functions can then be obtained by the following Legendre transformation
\begin{equation}
\Gamma [A^{\mu}_c]=\int d^3x J_{\mu}(x)\,A^{\mu}_c(x)- W[J]\,,
\end{equation} 
where
\begin{equation}
A_c^{\mu}(x)=\frac{\delta W[J]}{\delta J_{\mu}(x)}.
\end{equation}

Thus, the current-current correlation function that is needed for the Kubo formula can be obtained by taking the second derivative of the generating functional,
\begin{eqnarray}\label{correlation1sup2}
\langle j_\mu j_\nu \rangle=\frac{1}{e^2}\frac{\delta^2}{\delta A_c^{\mu} \delta A_c^{\nu}}\Gamma[A_c]\Big| _{A_c=0}\,.
\end{eqnarray}
It turns out, however, that the above expression is nothing but the $A_\mu$-field self-energy
$\Pi_{\mu\nu}$, also known as the vacuum polarization tensor, which is given by
 \begin{equation}
G_{\mu\nu}^{-1}-G_{0,\mu\nu}^{-1} = - e^2 \Pi_{\mu\nu}, 
\label{fermionSDEsup1}
\end{equation}
where $G$ is  the exact $A_\mu$-field Euclidean propagator and
$G_{0}$ is the free one,
\begin{equation}\label{fotonsup}
G_{0,\mu\nu}=\frac{1}{4\sqrt{p^2}}P_{\mu\nu}.
\end{equation}
In this expression, $P_{\mu\nu}=\delta_{\mu\nu}-p_\mu\, p_\nu/p^2$, and $p^2=p_3^2+\textbf{p}^2$, where $p_3$ is the third component of the energy-momentum vector in the Euclidean space.
We, therefore, come to the conclusion that
\begin{eqnarray}\label{correlation1sup}
\langle j_\mu j_\nu \rangle_{1PI}= \Pi_{\mu\nu}. 
\label{fc}
\end{eqnarray}

$\Pi^{\mu\nu}$ has been calculated 
up to the order of two-loops in PQED, for the case of two-component fermions. The Euclidean one-loop contribution for a single massless fermion, which is the
same for QED3, is \cite{Luschersup}
\begin{equation}
\Pi_{\mu\nu}^{(1)}(p)=A(p) \,P_{\mu\nu}+ B \,\epsilon_{\mu\nu\alpha}p^\alpha\,,
\label{eq1}
\end{equation}
where $A(p)=-\sqrt{p^2}/16$, and $ B =(1/2\pi) \left(n+ 1/2 \right)$,
with $n$ integer. Notice that the second term contains the P and T anomaly. Accordingly to the Coleman-Hill theorem \cite{colemanhill}, if the parity anomaly occurs, it will appear only in one loop fermion integration, i.e, only $\Pi_{\mu\nu}^{(1)}$ could have a term $\epsilon^{\mu\nu\alpha}p_{\alpha}$.

The two-loops contribution to the vacuum polarization tensor in PQED was calculated in Ref.\cite{Teber} and reads 
\begin{equation}\label{twoloops}
\Pi^{(2)}_{\mu\nu}= A(p)\, C_\alpha \alpha_g  P_{\mu\nu}, 
\end{equation}
with $C_\alpha = (92 - 9 \pi^2)/18\pi $. Indeed, there is no correction for the $B$-term.
Note that  $C_\alpha=0.056$ and the fine structure constant for graphene is $\alpha_g\sim 300/137=2.189$. It follows that  $C_\alpha \alpha_g <1$, thus justifying our perturbative calculation.

According to Eq.~(\ref{fc}), the irreducible current-current correlation function is given by 
\begin{equation}
\langle j^\mu j^\nu \rangle =j_1(p)\,P^{\mu\nu}+j_2\,\epsilon^{\mu\nu\alpha}p_{\alpha}\,,
\end{equation}
with 
\begin{equation} \label{j1sup}
j_1(p)=-N_f A(p)(1+C_\alpha \alpha_g+O(e^4)) 
\end{equation}
and
\begin{equation} \label{j2sup}
j_2=-N_f B, 
\end{equation}
where $N_f$ arose from the sum over all fermions.

\section{The dc conductivity}

Observe that the current correlator (\ref{correlation1sup}) is proportional to the number of flavors $N_f$, consisting of spin $\uparrow,
\downarrow$ and valleys $K, K^{\prime}$. We have, therefore $N_f = N_S + N_V$. 
The two spin components give identical contributions to Eq.~(\ref{kubosup}), therefore $N_S = 2$. 
We must be careful, however, when summing the contributions from the
two valleys $K$ and $K^{\prime}$. For symmetry reasons, it is reasonable to expect that both valleys will contribute identically.
Nevertheless, the valleys $K$ and $K^{\prime}$ are related to each other by TRS and, consequently their contribution
will depend on whether this symmetry is spontaneously broken or not.
When TRS symmetry is preserved, both valleys clearly give identical contributions and $N_V=2$ or $N_f=4$.

Indeed, in linear response theory, for each valley we have
\begin{equation}
\langle 0|j^i |0\rangle _K  = \frac{ i \langle 0| j^i_K  j^j_K |0\rangle} {\omega} A^j
\end{equation}
and
\begin{equation}
\langle 0|j^i |0\rangle_{K^{\prime}} = \frac{i \langle 0| j^i_{K^{\prime}} j^j_{K^{\prime}} |0\rangle}{\omega} A^j\,.
\end{equation}

The contribution from the two valleys to the average total current is given by  $\langle 0|j^i_K + j^i_{K^{\prime}}|0\rangle$.
According to the result above, this can be expressed as
\begin{eqnarray}
\langle 0|j^i |0\rangle_K  + \langle 0|j^i |0\rangle_{K^{\prime}} &=&\nonumber \\  \left\{\frac{ i \langle 0| j^i_K  j^j_K |0\rangle}{\omega}\right.
&+& \left. \frac{i\langle 0| j^i_K  j^j_K |0\rangle_T} {\omega}\right\} A^j\,.
\label{jkk1}
\end{eqnarray}

Therefore, when the TRS is not spontaneously broken,
the sum of the contributions from the two valleys to the conductivity is
\begin{equation}
\sigma^{ik}= \lim_{\omega\rightarrow 0, \textbf{p}\rightarrow 0}\left\{\frac{i \langle j^i j^k \rangle}{ \omega}
+ \frac{i \langle j^i j^k \rangle^T}{ \omega} \right\},
\label{kubo1sup}
\end{equation}
where $\langle j \ j \rangle^T$ is the time-reversed  correlator and the sum over
spins is assumed to have been done, namely, at this level $N_f = N_S =2$.

Now, observe that, according to Eqs.~(\ref{j1sup}) and (\ref{j2sup}), $j_2$ is a constant, whereas $j_1$ is a
function of $ \textbf{p}^2 + p_3^2 $ in Euclidean space. When we go back to the real time, we must analytically continue $p_3$ to the imaginary axis. Hence, $j_1$ becomes a function of
$ \textbf{p}^2 + (i\ p_0)^2 $. This is invariant under time-reversal ($i \rightarrow -i$ , $p_0 \rightarrow p_0$ and
$\textbf{p} \rightarrow -\textbf{p}$) and, consequently, so is $j_1$.

In the limit $\textbf{p}\rightarrow 0$, the current correlator and its time-reversed version are given, respectively, by
expressions of the form
\begin{equation}
\label{jjjsup}
\langle j^i j^k \rangle = j_1((i\ p_0)^2) \,\delta^{ik} + j_2  \epsilon^{ik0} (i\ p_0)\,,
\end{equation}
\n and
\begin{equation}
\langle j^i j^k \rangle^T = j_1((-i\ p_0)^2)\, \delta^{ik} + j_2  \epsilon^{ik0} (-i\ p_0)\,.
\label{jjjjsup}
\end{equation}

The first term is clearly  invariant, since $(i\ p_0)^2 =(-i\ p_0)^2$. The second term, conversely, is clearly non-invariant and derives from the anomalous part of the vacuum polarization tensor, which is generated by vacuum fluctuations. The $p_0$ variable above, in the unit system that we are using, must be identified  with the frequency $\omega$ in the Kubo's formula (\ref{kubosup}).

We may now take the limit $\omega \rightarrow 0$ , in order to get the dc conductivity.
It is worth mentioning that this limit in Kubo formula can be taken unambiguously when PQED is used to describe the
interactions, unlike the usual $\text{QED}_3$. This occurs due to the peculiar structure of the gauge field propagator of the theory, which produces a linear $\omega$-dependence in the current correlator for $\textbf{p}\rightarrow 0$, that will cancel the $\omega$ in the denominator in the Kubo's formula.

Using Eqs.~(\ref{jjjsup}) and (\ref{jjjjsup}), we see that the conductivity has the general form
\begin{equation}
\sigma^{ik}=\sigma^{xx}\delta^{ik}+\sigma^{xy} \epsilon^{ik0}.
\end{equation}

Inserting Eqs.~(\ref{jjjsup}) and (\ref{jjjjsup}) into Eq.~(\ref{kubo1sup}), we find that for an unbroken TRS phase only the 
longitudinal part survives. It is easy to see that we have equal contributions from the two valleys, hence $N_V=2$ or $N_f=4$.
 The $p_0$ dependence
cancels nicely and we can take the zero frequency limit without hurdles. Using Eq.~(\ref{j2sup}),  
 we obtain 
\begin{equation}
\sigma^{xx}= \left(\frac{\pi}{2} \frac{e^2}{h}\right)\left[1+\left(\frac{92 - 9 \pi^2}{18\pi} \right)\,\alpha_g 
+{\cal O}(e^4)\right] 
\end{equation}
and
\begin{equation}
\sigma^{xy}= 0\,.
\end{equation}

\section{Valley conductivity and the quantum valley Hall effect}

The average valley current is defined by
\begin{equation}
\langle J^i_V \rangle = \langle 0|j^i_K|0\rangle - \langle 0|j^i_{K^{\prime}}|0\rangle.
\label{marino}
\end{equation}
It vanishes whenever the two valleys contribute the same amount to the
electric current.
From Eq.~(\ref{marino}) it is clear that
\begin{eqnarray}
\langle J^i_V \rangle =  \left\{\frac{ i \langle 0| j^i_K  j^j_K |0\rangle}{\omega}\right.
- \left. \frac{i\langle 0| j^i_K  j^j_K |0\rangle_T} {\omega}\right\} A^j\,.
\label{jkk}
\end{eqnarray}

We can therefore define a dc ``valley conductivity'', which is given by
\begin{equation}
\sigma^{ik}_V= \lim_{\omega\rightarrow 0, \textbf{p}\rightarrow 0}\left\{\frac{i \langle j^i j^k \rangle}{ \omega}
- \frac{i \langle j^i j^k \rangle^T}{ \omega} \right\}.
\label{kubovsup}
\end{equation}

Now, using Eqs.~(\ref{jjjsup}) and (\ref{jjjjsup}) whith $N_f=N_S=2$ as before, one immediately concludes that the longitudinal parts
cancel, whereas the transverse component survives. The valley conductivity, therefore, is given by
\begin{equation}
\sigma^{xy}_V=4\left(n+\frac{1}{2}\right)\frac{e^2}{h},
\label{svxy}
\end{equation}
for $n=$ integer. The above result is exact, as a consequence of the Coleman-Hill theorem.

The longitudinal component, conversely, vanishes:
\begin{equation}
\sigma^{xx}_V= 0.
\end{equation}

The existence of a transverse valley conductivity characterizes the occurrence of a quantum valley Hall
effect. It is caused, ultimately, by the presence of the anomalous P,T violating term that appears in the
vacuum polarization tensor or, equivalently, in the current  correlator.

The anomalous terms are related to electron masses that are dynamically generated. These, however, arise in
pairs of opposite signs and, for an even number of flavors, cancel when summed, according to the Vafa-Witten theorem \cite{vw}.
There is, consequently no overall P, T violation and for this
reason the quantum Hall effect does not occur. The existence of individually violating terms, nevertheless, is
sufficient to produce a quantum valley Hall effect, which is analogous to the quantum spin Hall effect, but with spins replaced by valleys.

\section{Dynamically generated discrete energy states}

It was recently shown in Ref.~\cite{VLWJFsup} that a dynamical fermion mass generation occurs in PQED$_3$. In that context, where four-component massless fermions were used, the mass term breaks the chiral symmetry. For the two-component fermions considered here, however, the mass terms would break the parity symmetry instead. This non-symmetric phase may be investigated by considering non-perturbative solutions of the Schwinger-Dyson equation \cite{Appelquist2sup}, given by
\begin{equation}
S^{-1}_F(p)=S_{0F}^{-1}(p)-\Sigma (p), \label{fermionSDEsup}
\end{equation}
where $S_F(p)$ is  the full fermion propagator,
$S_{0F}(p)$ is the bare fermion propagator and $\Sigma(p)$ is the self-energy, which is given by
\begin{eqnarray}\label{fermionsup}
\Sigma (p)=\frac{e^2}{2}\rm{tr}\int\frac{d^3k}{(2\pi)^3} \gamma^{\mu}S_F(k)\gamma^{\nu}\, G_{\mu\nu}(p-k)\,,
\end{eqnarray}
where $G_{\mu\nu}$ is the full field propagator of the gauge field and $\rm{tr}$ is the trace over Dirac indexes.

By making a Taylor expansion around $\epsilon$
\begin{eqnarray}
\Sigma(p)=\Sigma(p=\epsilon)+(\gamma^\mu p_\mu-\epsilon)\frac{\partial\Sigma(p)}{\partial p}\huge|_{p=\epsilon}+...\end{eqnarray}
and imposing
\begin{equation}
\Sigma(p=\epsilon)=\epsilon,
\label{autosup}
\end{equation}
we may write the full fermion propagator as
\begin{eqnarray}
S_F(p)&=&\frac{1}{\gamma^\mu p_\mu-\Sigma(p)}\nonumber\\
&=&\frac{1}{(\gamma^\mu p_\mu-\epsilon)(1-\frac{\partial\Sigma(p)}{\partial p}\huge|_{p=\epsilon}+...)}
\nonumber\\
&=&\frac{\gamma^\mu p_\mu+\epsilon}{(p^2 - \epsilon^2)(1-\frac{\partial\Sigma(p)}{\partial p}\huge|_{p=\epsilon}+...)}.
\label{sfsup}
\end{eqnarray}

We see that  $\epsilon$ is the pole of the full physical electron propagator at zero momentum, being therefore
an eigen-energy. Using an $e^2$-expansion, the gauge field propagator can be written as 
\begin{equation}
G_{\mu\nu}\approx \frac{1}{\sqrt{p^2} (4+\frac{\lambda}{16})}\,P_{\mu\nu}, \label{propaxu}
\end{equation}
where $\lambda =e^2 N_f$.

Inserting Eq.~(\ref{fermionSDEsup}) and Eq.~(\ref{propaxu}) into Eq.~(\ref{fermionsup}), we obtain the integral equation
\begin{equation}
\label{sigmap}
\Sigma(p)=\frac{2\lambda}{N_f}\int\frac{d^3k}{(2\pi)^3}\,\frac{\Sigma(k)\,}{k^2+\Sigma^2(k)}\,\frac{1}{\sqrt{(p-k)^2}(4+\frac{\lambda}{16})}.
\end{equation}

Introducing an ultraviolet cutoff $\Lambda$, we can transform the integral equation (\ref{sigmap}) into a differential equation (Euler's equation),
\begin{equation}
\frac{d}{dp}\left(p^2\frac{d\Sigma(p)}{dp}\right)+\frac{N_c}{4 N_f}\Sigma(p)=0,
\end{equation}
where
\begin{equation}
N_c = \frac{4\lambda}{\pi^2 (4+\frac{\lambda}{16})} \label{Nc}
\end{equation}
is a critical number of flavors.
The self-energy also obeys
\begin{equation}\label{uvsup}
\lim_{p\rightarrow \Lambda}\left(2 \,p \,\frac{d\Sigma (p)}{dp}+\Sigma (p)\right)=0,
\end{equation}
and
\begin{equation}\label{irsup}
\lim_{p\rightarrow 0}p^2\,\frac{d\Sigma (p)}{dp}=0,
\end{equation}
representing the ultraviolet (UV) and infrared (IR) boundary conditions, respectively.

The solutions of Euler's differential equation are
\begin{equation}
\Sigma(p)={\tilde C} p^{a_{+}}+ {\tilde D} p^{a_{-}},
\label{EulerSol}
\end{equation}
where $a_{\pm}=-1/2\pm1/2\sqrt{1-N_c/N_f}$ and ${\tilde C}$ and ${\tilde D}$ are constants.
The solution Eq.(\ref{EulerSol}) can, without loss of generality, be rewritten as 
\begin{equation} 
\Sigma(p)=
\frac{(C + D)}{\sqrt{p}} \cos \left( \gamma \ln \frac{p}{\bar \Lambda} \right) + 
i \frac{(C - D)}{\sqrt{p}} \sin \left( \gamma \ln \frac{p}{\bar \Lambda} \right) ,
\label{Rewritten}
\end{equation}
where $C = {\tilde C} {\bar \Lambda}^{i \gamma}$, $D = {\tilde D} {\bar \Lambda}^{- i \gamma}$, and the constant 
\begin{equation}
\gamma=\frac{1}{2}\sqrt{\frac{N_c}{N_f}-1}
\end{equation} 
is real in the range of couplings such that $N_f<N_c$. 
Insertion of Eq.~(\ref{Rewritten}) into the boundary condition (\ref{uvsup}) provides
us with the constraints on the values of $C$ and $D$, namely
\begin{eqnarray}
\lim_{p\rightarrow \Lambda} \frac{ \gamma}{\sqrt{p^3}} \left[ (C+D)  \sin \left( \gamma \ln \frac{p}{\bar \Lambda} \right)  \right. \\ 
 \left. + i (D - C) \cos \left( \gamma \ln \frac{p}{\bar \Lambda} \right) \right] = 0.
\end{eqnarray}

There are two possible solutions that obey the constraint: either $C = D$ and $\sin \left[ \gamma \ln \left(\Lambda/ \bar \Lambda \right) \right] = 0$ 
or  $C = -D$ and $\cos \left[ \gamma \ln \left(\Lambda/\bar \Lambda \right) \right] = 0$. 

By assuming $2C = \Lambda^{3/2}$ to regularize the constraint, we can rewrite Eq.~(\ref{Rewritten}) as  
\begin{eqnarray} \label{solmsupa}
\Sigma(p)&=&\frac{\Lambda^{3/2}}{\sqrt{-p}}\sin\left(\gamma\ln{\frac{p}{\bar\Lambda}}\right), \quad C = -D \\
\Sigma(p)&=&\frac{\Lambda^{3/2}}{\sqrt{p}}\cos\left(\gamma\ln{\frac{p}{\bar\Lambda}}\right), \quad C = D
\label{solmsupb}
\end{eqnarray}
where
the constant $\bar\Lambda$ can be obtained from Eq.~(\ref{uvsup}), namely
\begin{eqnarray}
\bar\Lambda &=& \Lambda\,\exp\left[-\frac{(2 l + 1) \pi}{2 \gamma} \right], \quad C = -D, \\
\bar\Lambda &=& \Lambda\,\exp\left[-\frac{ k \pi}{ \gamma} \right], \quad C = D, 
\end{eqnarray}
with $k$ and $l$ integers. Now, we chose these integers to have a value as small as possible, but in a way to guarantee that $ \Lambda \ge \bar\Lambda$.
This choice fixes $l = k = 0$.

In order to obtain the physical eigenenergies $\epsilon$, we must solve Eq.~(\ref{autosup}). Using
the Eqs.~(\ref{solmsupa}) and (\ref{solmsupb})  for the self-energy, we have
\begin{eqnarray}\label{2solmsup}
\epsilon&=& -\frac{\Lambda^{3/2}}{\sqrt{-\epsilon}}  \cos\left(\gamma\ln{\frac{|\epsilon|}{\Lambda}}\right), \quad \epsilon < 0, \\
\epsilon&=&\frac{\Lambda^{3/2}}{\sqrt{\epsilon}}  \cos\left(\gamma\ln{\frac{|\epsilon|}{\Lambda}}\right), \quad \epsilon > 0.
\label{1solmsup}
\end{eqnarray}

Let us now define
\begin{eqnarray}
 z = -\gamma \ln\left(|\epsilon|/\Lambda\right),\,\,\,\,\,\,(z>0)
\label{x}
\end{eqnarray}
or, equivalently,
\begin{equation}
|\epsilon| = \Lambda\  \exp\left(- \frac{z}{\gamma}\right).
\label{2sup}
\end{equation}

Then, after inserting Eq.~(\ref{2sup}) into Eqs.~(\ref{2solmsup}) and (\ref{1solmsup}), we find that
the dimensionless quantity $z(\gamma)$ is given by the solutions of the transcendental equation
\begin{equation}
\exp\left(-\frac{3z}{2\gamma}\right) =  \cos z, 
\label{3sup}
\end{equation}
which holds for both, $\epsilon$ positive or negative. Its solutions depend on $\gamma$, which on its turn is determined by the coupling and the number of flavors
$N_f$.

Let us call $z_n = Z_n$, $n=0, 1, 2, ...$ the solutions of Eq.~(\ref{3sup}).
It is not difficult to infer, from the graphic representation of the
functions in Eq.~(\ref{3sup}), that
\begin{eqnarray}
Z_n &=& n \pi+  \delta_n
\end{eqnarray}
where
\begin{eqnarray}
&\ &0 \leq \delta_n \leq \frac{\pi}{2} \ \ \ \ \  n=0, 2, 4, ... \nonumber \\
&\ & \frac{\pi}{2} \leq \delta_n \leq \pi \ \ \ \ \  n=1, 3, 5, ...
\end{eqnarray}

For all values of $n$, $\delta_n \rightarrow \pi/2 $ for $n \rightarrow \infty$
whereas  $\delta_n \rightarrow 0 $ ($n$ even) and $\delta_n \rightarrow\pi$ ($n$ odd) for $\gamma \rightarrow \infty$
(an unphysical limit, in which $N_f \rightarrow 0$ ).

The energy levels are then
\begin{eqnarray}
\epsilon^{(\pm)}_n&=& \pm\Lambda\, \exp\left\{-\frac{Z_n}{\gamma}\right\} \nonumber \\
&=& \pm \Lambda\, \exp\left\{-\frac{1}{\gamma}\left( n \pi +  \delta_n \right)\right\}.
\end{eqnarray}

Observe that the negative energy levels increase and the positive ones decrease with $n=0, 1, 2, ...$, in such a way that both of them accumulate in
zero for $n \rightarrow \infty$. In the situation when $\gamma \rightarrow 0$, which
occurs when $N_c \rightarrow N_f$, all energy levels $\epsilon^{(\pm)}_n$ collapse to zero, thus
destroying the effect. Therefore, $N_c$  is a critical quantity for the phenomenon
we described: this will only occur for  $N_c > N_f$ or equivalently for $\lambda>\lambda_c$ (see Eq.~(\ref{Nc})$\,)$.  Since $\gamma$ is small, the LHS of the Eq.~(\ref{3sup}) tends to zero and the solutions are the zeros of the cosine function, which are $Z_n \approx
(2 n + 1 ) \pi /2$.  

Since the eigenenergies are zero-momentum poles of the corrected electron propagator, they become dynamically generated electron masses. Notice that
all flavors will acquire a mass
\begin{equation}
M = \epsilon^{(\pm)}_0 \simeq \pm\Lambda\, \exp\left\{-\frac{\pi}{2 \gamma}\right\}.
\label{massa}
\end{equation}

We may now estimate the maximal energy gap, namely, $\Delta_\infty = \epsilon^{(+)}_0-\epsilon_\infty=\epsilon_\infty-\epsilon^{(-)}_0 $.
This is given by
\begin{eqnarray}
\Delta_\infty = |\epsilon^{(\pm)}_0 | \simeq
\Lambda\, \exp\left\{-\frac{\pi}{2 \gamma}\right\},
\end{eqnarray}
where we used the fact that $\delta_0 \simeq \pi/2$. The first gap is $\Delta_1=|\epsilon^{(\pm)}_0|-|\epsilon^{(\pm)}_1| \approx \Delta_\infty $, since $\gamma$ is small.

The energy gap $\Delta_\infty$ determines the critical temperature $T_c$ for observing the quantum valley Hall conductivity that we have
predicted: for $T>T_c$ the plateaus will be washed out by thermal activation.
Assuming that a fraction $1-x$ of the electrons in the ground state are promoted to higher levels by thermal activation
($N/N_0 = 1-x$ ), we have
\begin{equation}
N= N_0 \exp\left[-\frac{\Delta_\infty}{k_B T}\right],
\end{equation}
or equivalently
\begin{equation}
T_c= \frac{\Delta_\infty}{k_B \ln (N_0/N)} \simeq \frac{\Delta_\infty}{k_B x}.
\end{equation}

We find, therefore
\begin{equation}
T_c \simeq \frac{\Lambda}{k_B x}\exp\left[-\frac{\pi}{2 \gamma}\right].
\end{equation}
Notice that $k_B T_c$ is of the order of the modulus of the electron masses and hence $T_c\sim 0.05\,K$. 

It is worth to emphasize that the study of dynamical mass generation for electrons in graphene has been investigated in the literature in different contexts, by considering only the static Coulomb interaction with screening effect. In this case, the quantum corrections in the gauge propagator contain only the ``00" component of the vacuum polarization \cite{khveshchenko,gorbar,khveshchenko1}.  The influence of the renormalization of the Fermi velocity for the gap equation was investigated in reference \cite{khveshchenko1}.

We will see in the next section that the fact that there are positive and negative masses is crucial for guaranteeing that the
system obeys the Vafa-Witten theorem, although individually each valley breaks 
the P and T symmetries. 

\section{The Vafa-Witten Theorem}

Here, we examine our results in the light of the Vafa-Witten theorem \cite{vw}.
Let us start by reviewing the proof of the theorem.
Consider the partition function of a gauge field with a vector minimal coupling to a Dirac field of mass
$M$ in Euclidean space.
Integration over the Dirac field yields
\begin{eqnarray}
{\cal Z}[\xi]&=&{\cal N}\int DA_\mu  \, {\rm e}^{- S_{\xi}[A_\mu]  } Det[D\!\!\!\!/ +M][A_\mu].
\label{a}
\end{eqnarray}

On general grounds, the bosonic part of the action may be decomposed into a P (and T) invariant part $S_0$ and a
P (and T) non-invariant part $i \xi X$, which in Euclidean space is purely imaginary. Indeed,
in the expression above $ S_{\xi} = S_0  + i \xi X$ \cite{vw}, where $\xi$ is a real parameter.

The theorem follows from the fact that, for a real and positive  fermionic determinant,  evidently, we have a bound
${\cal Z}[\xi] \leq {\cal Z}[0]$. Defining  ${\cal Z}[\xi] = e^{-{\cal V}[\xi]}$, we have ${\cal V}[\xi] \geq {\cal V}[0]$.
 Hence, the energetically most stable state is the one with $\xi=0$, implying $\langle X \rangle =0$, which means
that there is no spontaneous breakdown of P and T symmetries. This completes the proof.

A key ingredient for the demonstration of the Vafa-Witten theorem is the fact that the fermionic determinant must be real and positive. This is guaranteed  by the following lemma:
Suppose the anti-hermitian operator $D\!\!\!\!/$ has eigenvalues $i\lambda$; 
\begin{eqnarray}
\left [D\!\!\!\!/ + M\right ] \psi  = \left [ M + i \lambda\right ] \psi .
\label{b}
\end{eqnarray}

Since the $\gamma^5$-matrix anticommutes with $D\!\!\!\!/$, it follows that for each eigenstate $\psi $ there will be another
one given by $\gamma^5 \psi $, with eigenvalue $ \left [ M - i \lambda\right ] $. The fermionic
determinant, accordingly will be $\prod_\lambda [ M^2 + \lambda^2]$, which is real and positive, thus completing
the proof of the lemma.

We now  come to the system we are using for describing graphene. A great difference with respect to
the framework where the Vafa-Witten theorem has been demonstrated is the fact that there is no
$\gamma^5$-matrix for two-component Dirac fermions in two spatial dimensions, hence the above lemma,
which forced the fermionic determinant to be real and positive, does not apply.

The fermionic determinant was actually calculated in
Ref.~\cite{Luschersup} for a single two-component fermion in two-dimensional space  and indeed, it presents a complex phase. This
is proportional to a Chern-Simons term, which
is not invariant either under P or T, and the proportionality factor is fixed and non-vanishing. In this case the
theorem clearly does not apply. The bound on the partition function just cannot be fulfilled.

Now consider the case of many-flavor fermions. Then, we have the product of all flavor determinants, which results in a
real positive modulus plus an overall phase given by the sum of the complex phases
of all flavors. 
For fermions of mass $M$, each phase is proportional to $M/|M|$, namely,
to the mass' sign. This fact leads us to conclude, by using the same argument employed in the
demonstration of the theorem, that the anomalous phases would cancel for an even
number of flavors provided there is the same number of masses with opposite signs.
This would make the resulting many-flavor determinant real and positive and would redeem the result
 of P and T invariance.

In our system, specifically, we have just seen in the
previous section that in the low-temperature phase
the dynamically generated electron masses present two opposite signs: $M = \pm |\epsilon_0|$, hence the anomalous complex phases will cancel in compliance with the Vafa-Witten theorem.

The dynamical generation of masses and the associated occurrence of complex phases in the fermionic determinants, even though
cancelling when fully summed, are responsible for the onset of a non-vanishing valley current below Tc, which
characterizes a Quantum Valley Hall effect.

This is the ``center-of-gravity'' of this work. The point where
the dynamical generation of electron masses, obtained from the electron self-energy, meets the dynamical generation
of a P and T violating term in the vacuum polarization, for each flavor. By summing over the even number of flavors, the anomalous
terms do cancel as a consequence of the fact that the masses are generated in pairs of opposite signs. This form of
mass generation, despite ruling out a regular Quantum Hall effect, however does
imply a Quantum Valley Hall effect.

\section{Fermi Velocity}

For $v_F\neq c$, we must rewrite the electronic kinetic term and the current, respectively as
\begin{equation}
i\partial\!\!\!/ =i\gamma^0\partial_0+i\,v_F\gamma^i\partial_i,
\end{equation}
and
\begin{equation}
j^\mu=e\,\bar\psi\gamma^\mu\psi=e\,(\bar\psi\gamma^0\psi,v_F \,\bar\psi\gamma^i\psi).
\end{equation}

Now, when evaluating the current correlation function given by Eqs.~(\ref{correlation1sup}-\ref{twoloops}), we must 
 replace $\gamma_i\rightarrow v_F\,\gamma_i$ in the vertices.

The one-loop result in momentum space is
\begin{equation}
\Pi^{00}(p_0,\textbf{p})=-\frac{1}{16}\frac{\textbf{p}^2}{\sqrt{v^2_F\textbf{p}^2+p_0^2}},
\end{equation}
\begin{eqnarray}
\Pi^{i0}(p_0,\textbf{p})&=&\frac{1}{16}\frac{p^0\,p^i}{\sqrt{v^2_F\textbf{p}^2+p_0^2}}\nonumber \\
&+&\frac{1}{2\pi}\left(n+\frac{1}{2}\right)\,\epsilon^{i0j}\,p_j\,,
\end{eqnarray}
and
\begin{eqnarray}
\Pi^{ij}(p_0,\textbf{p})&=&-\frac{1}{16}\left[\frac{\delta^{ij}\left(v^2_F\textbf{p}^2+p_0^2\right)-v^2_F p^i\,p^j}{\sqrt{v^2_F\textbf{p}^2+p_0^2}}\right]\nonumber \\
&+&\frac{1}{2\pi} \left(n+\frac{1}{2}\right)\,\epsilon^{ij0}\,p_0\,.
\end{eqnarray}

The generating functional ${\cal Z}[J]$ is obtained by performing different
gaussian integrals over $A_0$ and $A_i$. Then, it is easy to see from Eq.~(\ref{correlation1sup}) that
the current correlator will be expressed in terms of
$\Pi_{ij}$, $\Pi_{00}$ and $\Pi_{i0}$. After taking the limit ${\bf p} \rightarrow 0$ in the
Kubo formula, we conclude that  the only contribution comes from $\Pi_{ij}$, since $\Pi_{00}$ and $\Pi_{i0}$
vanish in this limit. From the equation above, however, we see that all dependence on $v_F$ disappears
in the limit when the external momentum ${\bf p} \rightarrow 0$. Note that the above argument also holds for the two-loops contribution. 

It is quite interesting to note that all dependence  on $v_F$ coming from the momentum integration of the internal
fermion lines is completely removed by a scale transformation, which is possible because the fermions are massless. This explains why all $v_F$ dependence comes from the external lines.

Let us consider now the effects of $v_F \neq 1$ and $c \neq 1$ on the self-energy $\Sigma(p)$, on the dynamically generated gaps $\epsilon_n$, and on the transition temperature $T_c$. Now, the previous argument with the scale transformation cannot be used despite the fact that the propagators are massless, because each of them  contains a different velocity.

In the relativistic case ($v_F = c =1$) the self-energy is a function $\Sigma(\sqrt{p_0^2+\textbf{p}^2})$. When we reinstate the physical values of $v_F$ and $c$, it happens that the self-energy becomes a function $ \Sigma(f_1(v_F,c)p_0,\, f_2(v_F,c)\textbf{p})$ \cite{nagaosa}, where the coefficient $f_1(v_F,c)$ is dimensionless, whereas $f_2(v_F,c)$ has dimension of velocity. Note that $p_0$  has dimension of energy when we use the physical units.

We are interested in the dynamically generated gap, i.e., the mass spectrum; hence, we only need to evaluate $\Sigma(f_1\, p_0, f_2\,\textbf{p} = 0)$. Therefore, we make the Taylor expansion of the self-energy in the variable $p_0$ around the gap $ \epsilon$, namely
\begin{eqnarray}
\Sigma(f_1 p_0)=\Sigma(f_1 \epsilon)+(\gamma^0 p_0-\epsilon)\frac{\partial\Sigma(f_1 p_0)}{\partial p_0}\huge|_{p_0=\epsilon}+...\end{eqnarray}

Now, we must impose the condition
\begin{equation}
\Sigma(f_1 \epsilon)=\epsilon,
\label{Neweq}
\end{equation}
instead of (\ref{autosup}).

The full fermion propagator at zero momentum becomes
\begin{eqnarray}
S_F(p_0,\textbf{p}=0)&=&\frac{1}{\gamma^0 p_0-\Sigma(f_1 p_0)}\nonumber\\
&=&\frac{1}{(\gamma^0 p_0-\epsilon)(1-\frac{\partial\Sigma(f_1p_0)}{\partial p_0}\huge|_{p_0=\epsilon}+...)}
\nonumber\\
&=&\frac{\gamma^0 p_0+\epsilon}{(p_0^2 - \epsilon^2)(1-\frac{\partial\Sigma(f_1 p_0)}{\partial p_0}\huge|_{p_0=\epsilon}+...)}
\label{} \nonumber
\end{eqnarray}
and the  dynamically generated gap is still $\epsilon$.
This is determined by Eq.~(\ref{Neweq}), which yields the solutions
\begin{eqnarray}
\bar{\epsilon}^{(\pm)}_n&=&\pm\Lambda\, \exp\left\{-\frac{\bar{Z}_n}{\gamma}\right\}, 
\end{eqnarray}
where $\bar{Z}_n$ are solutions of the equation
\begin{equation}
\exp\left(-\frac{3}{2\gamma}z \right) = f_1(v_F,c)\cos z.
\label{net}
\end{equation}

For physical values of the coupling constant of graphene, $\gamma$ is rather small. It follows that the left hand side of Eq.(\ref{net}) is close to zero, as before. Consequently, the solutions of Eq.(\ref{net}) are effectively given by the zeros of the cosine function, independently of the value of $f_1(v_F,c)$. Hence, we conclude that  $ \bar{Z}_n$ coincide with $Z_n$ and the dynamically generated gaps $\bar{\epsilon}^{(\pm)}_n$ are the same as before. This fact implies that our estimate for the transition temperature $T_c$ remains unchanged when the physical  values of $v_F$ and $c$ are used.

\section{Non-relativistic limit}

Now, let us investigate the small $v_F/c$ limit of the Dirac equation, assuming we are in the phase where the  energy states $\epsilon^{(\pm)}_n$ are present and give a mass to the electrons. Then, the Foldy-Wouthuysen transformation can be applied to the Dirac equation coupled to the pseudo-electromagnetic field. This result can be easily obtained from the corresponding transformation in QED$_4$ \cite{Bjorkensup}, simply by constraining the matter to move only in the $x-y$ plane  with $J_z=0$ (no current matter in the $z$ direction).

Using the Fermi velocity divided by the light velocity as an expansion parameter, the non-relativistic limit of the Dirac equation in the lowest approximation yields the Pauli equation, which contains: (a) the minimal coupling with the vector potential $\propto ({\bf p}-{\bf A})^2$, (b) an electron-spin interaction with the magnetic field $\propto ({\bf \sigma} \cdot {\bf B})$, and (c) the static Coulomb interaction $\propto (1/r)$.

 In the absence of a magnetic field, the second order term in the expansion  gives other interactions related to the electric field: (a) a Darwin interaction $\propto \rho({\bf r})$ and (b) a  spin-orbit term which, taking $J_z=0$, reduces to a Rashba-like spin-orbit coupling. By applying an electric field in the $z$-direction, for instance, we obtain a spin-orbit coupling $\propto (\sigma_xp_y-\sigma_yp_x)$. It was recently shown that it is possible to generate quantum Hall states in the presence of a Rashba spin-orbit coupling and static interactions \cite{Beugelingsup,Shengsup}. Since the spin-orbit coupling is included in the full electromagnetic interaction and this produces the QVHE, there could be a relation between the two effects. We shall explore this connection elsewhere.

\section{Unitarity}

Here, we study the properties of the gauge field propagator in PQED, given by Eq.~(\ref{fotonsup}) and show, in particular,
that unitarity is preserved.
 
First of all, let us observe that this propagator has no poles, just a cut, hence the pure gauge field of PQED 
has no particle content, as should be expected. Evidently there are no photons in two-dimensional space. 
Pure PQED gauge theory, however is the bosonized version of the free massless Dirac field in 2D \cite{marinosup2}. It
has no particle content itself, in the same way as its counterpart in 1D, namely, the massless scalar field, which
is the bosonic field associated to the massless Dirac field in this case.

As usual,
we use the Feynman prescription $p^2 \rightarrow p^2 + i \epsilon$ $(p^2 = \omega^2 - \textbf{p}^2)$,
in order to define the Feynman propagator
\begin{equation}
G^{\mu\nu}_{0,F}(t, \textbf{x})  =\frac{1}{4} P^{\mu\nu} D_F(t, \textbf{x}) ,
\label{m0}
\end{equation}
where $P^{\mu\nu}$ is the transverse projector and $ D_F(t, \textbf{x}) $ is the  corresponding scalar propagator, namely
\begin{equation}
D_F(t, \textbf{x})  = \int \frac{d \omega}{2\pi} \int \frac{d^2k}{(2\pi)^2} 
\frac{e^{-i\omega t}\ e^{i\textbf{k}\centerdot \textbf{x}}}{[\omega^2 - \textbf{k}^2+ i \varepsilon]^{1/2}}.
\label{m1}
\end{equation}

This integral has been carefully calculated in Ref.~\cite{marinorubenssup} (see Appendix 1 therein). The result is
\begin{equation}
D_F(t, \textbf{x})  = \frac{C}{[t^2-r^2- i \varepsilon]},
\label{m2}
\end{equation}
where $r = |\textbf{x}|$ and $C=- 1/2\pi^2$.

Unitarity of the $S=1+i T$ operator, i.e, $S^{\dagger}S=1$, implies
\begin{equation}
i (T-T^{\dagger})=T^{\dagger}T,  \label{unicond}
\end{equation}

We first consider the scalar field case.
Considering the amplitude corresponding to the previous operator equation evaluated between states
 $|i\rangle$ and $|f\rangle$ and introducing a complete set of intermediate states $|x\rangle$ on the RHS, the above unitarity condition becomes
\begin{equation}
D_{if}-D^*_{if}= -i \sum_x  \int d\Pi \, (2\pi)^3 \delta^3(p_i-p_x)\, (D^*_{ix}D_{xf}),
\end{equation}
where  $D_{if}$ is given by $\langle f|T|i\rangle= (2\pi)^3 \delta^{3}(p_i-p_f) D_{if}$ and $d\Pi$ is the phase space factor, which is needed to ensure
that the sum over the intermediate states corresponds to the identity. The equation above is known as the generalized optical theorem. In the limit $i \rightarrow f$, we have that the Feymann propagator of the scalar field reads 
\begin{equation}
D_{ii}=D_F(\omega,\textbf{k})=\frac{1}{[\omega^2 - \textbf{k}^2+ i \varepsilon]^{1/2}},
\end{equation}
and obeys the equation
\begin{equation}
D_{ii}-D^*_{ii}=-i \sum_x  \int d\Pi \,  (2\pi)^3 \delta^3(0) (D^{*}_{ix}D_{xi}).  \label{unicond2}
\end{equation}
or
\begin{eqnarray}
D_F(\omega,\textbf{k})-D^*_F(\omega,\textbf{k})&=&
\nonumber \\
-i   \int d\Pi \,  (2\pi)^3 \delta^3(0) \int \frac{d \omega_x}{2\pi} \int \frac{d^2k_x}{(2\pi)^2}
\nonumber \\
D^*_F(\omega_x,\textbf{k}_x)D_F(\omega-\omega_x,\textbf{k}-\textbf{k}_x).  \label{unicond3}
\end{eqnarray}
We now Fourier transform the above equation back to coordinate space and, using Eq.~(\ref{m2}) and the fact that the
Fourier transform of a convolution is a product, we obtain
\begin{equation}
i C \frac{\varepsilon}{[t^2-r^2]^2 + \varepsilon^2} =-i C^2  \int d\Pi \,  \delta^3(0)
 \frac{1}{[t^2-r^2]^2 + \varepsilon^2}. 
\label{mm2}
\end{equation}
The phase space  integral above yields $ \tau / L^2$ \cite{peskin}, whereas $\delta^3(0) = \tau L^2$, 
where $L$ and $\tau$ are the characteristic length and time of the system, respectively. By choosing $\varepsilon = -C \tau^2= \tau^2/ 2\pi^2$ 
(notice that $\varepsilon$ has dimension of $\tau^2$), the two sides above coincide and unitarity is, therefore,
demonstrated. The corresponding result for the vector field $A_\mu$ follows by making $C \rightarrow C/4$
and from the transverse projector property: $P^2 = P$.

\end{document}